# A Model using agile methodologies for defining metrics to be used by the Public Sector in Brazil to set remuneration for outsourced software development

Washington Henrique Carvalho Almeida
Center of Advanced Studies and Systems of Recife
Recife, Brazil
whca@cesar.school

*Abstract*— The process of contracting software factories within the scope of the Federal Public Administration (APF, in Portuguese) in Brazil has undergone some changes due to legislative alterations and a model has been proposed to improve the rendering of services regarding delivering results. Software factory contracts based on predictive processes and on the metrics of Function Points are the target of criticisms and have issues in achieving results, as verified in previous research. The initial objective of the study now proposed will be to define a process model for formulating metrics that can be used in agile contracts as opposed to those of standardized Function Points which have already proved to be quite problematic and difficult for the contractor to manage. Thus, in line with the theme of agile contracts for software development companies, the study proposed will looking for better understand the problem of software development and its maintenance by public agencies by means of contracts using agile methodologies and appropriate metrics for remunerating these services.

I. PROBLEM

The Federal Public Administration (APF, in Portuguese) in Brazil outsources software development activities on a large scale, as it is well understood that Information and Communication Technology (ICT) is essential if the public sector is to offer quality services. The investments made in contracting ICT services must be undertaken with caution, be planned and take into consideration the amount of public financial resources available.

Therefore, how outsources software development are contracted has been the target of audits by the Tribunal of Accounts of the Union (TCU, in Portuguese) and by the Comptroller General of the Union (CGU, in Portuguese), these institutions have legal powers to investigate illegalities in Brazil [1]. The problem experienced in Brazil is largely due to culture and a heavy bureaucracy of legislation, where payment must be restricted to tangible deliveries, which is not always easy to highlight in software development that is an intellectual work.

In the research [1] found that there is a difficulty in delivering results and managing contracts due to the way in which services rendered are remunerated. The Function Points (FP) metric is the most widely used metric for assessing the size of software among software companies in Brazil. It should be emphasized that, in Brazil, there are norms [1] that indicate that the government and similar public bodies are obliged to use functional metrics of size when drawing up contracts.

Some public bodies have reported difficulties in adopting the FP metric due to e.g., the need to hire a company for FP count and possible different interpretations in the count. In addition, it was found that adopting this metric led to problems for public bodies regarding delivering results, as delivery delay and lack of quality this, was evidenced in the research [1]. Then, it was noted that, after a cycle of unsuccessful contracts, public bodies with more advanced levels of maturity started to adopt other measurement metrics and thereafter, some results began to appear, which pointed towards there being a new model for drawing up contracts.

Therefore, the question is raised as to whether the FP metric is used due to its relevance or only because it is recommended by the control public bodies and published standards. Examples of disadvantages cited in [2] included that the FP metric may not adequately remunerate the effort required to do the work; the measurement is not accurate, and includes neither architectural nor non-functional requirements, in addition to which it is not entirely suitable for the agile methodology.

One of the problems of adopting other metrics to replace the FP metric is that in general, as found in [1], effort metrics such as the Unit of Technical Services (UTS) require a service catalog to be written for later use and that the contracted party complies with it. The last time that the TCU evaluated the use of the UTS by Agreement 1508/2020, deficiencies were verified in estimating price and dimensioning quantities in weakly separated parameters, since this metric is constructed on a historical basis and uses an estimate based on analogy (ABE). Another one is Technical Service Hour (TSH), like UTS, but without the definition of a fixed catalog. The real impact of this prohibition is that there is no metric accepted to remunerate services and what creates an impasse in contracting these services.

In addition to these metrics, there are other variations for calculating the remuneration of services, e.g. based on the hour of work, but attached to deliveries [1]. However, there is no consensus on adopting a specific metric and there is no guarantee of a calculation that results in an adequate remuneration that works for all cases, or at least for most of them. Hundreds of metrics have been proposed for computer programs, but not all of them are practical [3]. Thus, on if adopting a metric suitable for the context is a preponderant factor for the success of an outsourcing contract, a new metric could be suggested that considers factors that result in a more adequate measurement for remunerating services when contracting ICT services for the APF, making it possible to adapt to the context of each organization, featuring a gap to be filled.

Once again, the problem is not in the metric, but in its bad construction, not considering the current legislation, besides the almost general deficiency of historical bases in public entities in Brazil.

From the above, the idea for a solution is to develop a process model for contracting ICT services and create a metric for calculate the remuneration that each institution may be

able to implement within a context. Therefore, based on these reasons, the research question will be:

**RQ:** *how to contract software development and maintenance services using agile methods and remunerating these contracts with a metric of low complexity for management due remuneration for the work performed?*

## II. RESEARCH HYPOTHESIS

Based on the question raised in the previous section, the following hypothesis is considered:

> *- It is possible to design a process model for defining a metric that remunerates by results the effort spent on rendering the service and that this process can be adapted for public bodies that decide to contract systems development and maintenance services.*

With the initial hypothesis, it was defined the methods that will be used in the following section.

## III. METHODS AND PRELIMINARY FINDINGS

Initially, an exploratory survey [4] was conducted, the objectives of which are to collect data, such as the main research bases, and to identify themes that will be used to guide [5] a Systematic Literature Review. Then, a Systematic Literature Review [6] was conducted with a view to reevaluating a relatively broad topic, and to identifying, analyzing, and structuring the objectives, methods and content of the primary studies conducted.

From the data collected in the systematic review [6], a process will be designed so that public bodies can create a metric consistent with the real effort made to render the software development service and so that this process can become a replicable model that other institutions can adopt, and also, in this context, so that a roadmap of premises can be adopted in contracts that use agile methods.

The literature shows metrics used in the most diverse contexts [7] [8] [9] [10]. From a previous systematic study [6], several metrics were found to assess the effort undertaken in constructing a functionality from a parametric model such as COCOMO or its subsequent version, COCOMO II, which requires a broad historical base, often drawing on functional measurement metrics such as FP and COSMIC. In fact, some studies also use LOC [11].

A metric can be used in compliance with the Brazilian normative framework for supplier remuneration whenever an APF body outsources software development, such as functional measurement metrics e.g., FP [12] and COSMIC [13]. In addition, other metrics, in more comprehensive studies, have been found, which can be applied to agile methods such as [14], Sprint Points [15], Story Points [16] and Delivery Stories [17]. In some cases, these have been combined with multi-factor techniques to improve accuracy and algorithms with a checklist, and even use machine learning [18]. Metrics are used in agile methods and measure effort, deadlines, and the costs and sizing involved in building software. Studies discuss agile methods and measure these 3 efforts, deadlines, costs aspects, and they also focus on software maintenance and bug fixing activities [16].

Some studies focus on using techniques based on learning with the most diverse techniques, from genetic algorithms, Bayesian statistics, fuzzy logic, neural networks, and machine learning with multiple approaches. But the ones most used are the techniques based on expertise with analogy (Expert-Based). Several uses of the metrics in agile methodologies are presented, which mix functional measurements such as FP or COSMIC, and classic agile metrics which are combined with multiple factors for calibrating accuracy [9].

The use of the FP Metric in the remuneration of services in contracts outside Brazil is identified. For example, the Italian public sector uses FP when outsourcing critical services, but the metric is used to assess the functional size derived from this productivity, with effort and cost. In addition, this study cites Scrum Points that are said to be a fixed number of hours within a sprint, for example 40 hours, and the number of deliveries made within this open scope system [7].

## IV. EXPECTED CONTRIBUTION

The study regards to the remuneration that is offered to a software development and maintenance in a contract, or for the development and maintenance of systems. It is important to better meet the needs of the contractor and the adequate remuneration of the contracted party, thereby seeking balance and harmony for a better rendering of services and a delivery of value. The adoption of metrics that do not directly associate the effort involved with the cost brings to discrepancies that lead to poor service delivery and consequently dissatisfaction of the contractors with the software factory model. Even with the use of alternative metrics to the metric FP, such as UTS, problems were also identified through an audit by TCU.

Therefore, the tangible output is an adaptable process model for software development and maintenance contracts, capable of subsidizing the creation of metrics for remuneration of effort through results within a context. The definition of a process and metrics model will guide the bodies in conducting hires most likely to be successful, this will be the main result of this research.

## V. EVALUATION PROCEDURES

The study will evaluate some metrics that have been used for measuring and estimating software projects and test some hypotheses regarding the contract model that should be adopted and assess which metric is the most appropriate. The research will be qualitative and will use the Case Study method, to support a validation of the model built for the adoption of the new contract model. A Focus Group will be formed to build the model, bringing in experts in the subject and representatives from public bodies with notable expertise in contracting software factories.

This evaluation methodology is a research technique that collects data by observing and recording the group's interaction with a topic determined by the researcher, with a view to obtaining personal perceptions of the members of the group on a defined area [19]. As a case study will be adopted, one of the problems is the lack of control of the environment in which it will be conducted. Thus, in this method, the behavior that will be implemented with the process for building the metrics will be described.

A preliminary plan is to design a process so that each public body can build its metrics for software development and maintenance contracts. To support the study better, the systematic review [6] discusses how metrics are used outside Brazil and of how they are applied in agile methodologies. The changes resulting from the change in labor legislation and the new law on outsourcing are also in the roadmap of the study.